\documentclass[12pt,preprint]{aastex}
\usepackage{amsmath}

\shorttitle{Main-Belt Comet P/2010 A2}
\shortauthors{Moreno et al.}

\begin{document}


\title{Water-ice driven activity on Main-Belt Comet P/2010 A2 (LINEAR)?}


\author{F. Moreno\affil{Instituto de Astrof\'\i sica de Andaluc\'\i a, CSIC,
  Glorieta de la Astronom\'\i a s/n, 18008 Granada, Spain}
\email{fernando@iaa.es}}

\author{
J. Licandro\affil{Instituto de Astrof\'\i sica de Canarias,
  c/V\'{\i}a 
L\'actea s/n, 38200 La Laguna, Tenerife, Spain, and 
 Departamento de Astrof\'{\i}sica, Universidad de
  La Laguna, E-38205 La Laguna, Tenerife, Spain}}  

\author{
G.-P. Tozzi\affil{INAF - Osservatorio Astrofisico di Arcetri, Largo E. Fermi
  5, I-50125 Firenze, Italy}}

\author{
J.L. Ortiz\affil{Instituto de Astrof\'\i sica 
de Andaluc\'\i a, CSIC, Glorieta de la Astronom\'\i a s/n, 
18008 Granada, Spain} }

\author{
A. Cabrera-Lavers\affil{
GTC Project Office, E-38205 La Laguna, Tenerife, Spain, and Instituto
de Astrof\'\i sica de Canarias, c/V\'{\i}a L\'actea s/n, 38200 La
Laguna, Tenerife, 
Spain}}

\author{
T. Augusteijn\affil{Nordic Optical Telescope, Apdo. de Correos 474,
  38700 Santa Cruz de la Palma, Spain}}

\author{
T. Liimets\affil{Nordic Optical Telescope, Apdo. de Correos 474,
  38700 Santa Cruz de la Palma, Spain, and Tartu Observatory 61602, 
T\~oravere, Tartumaa, 
Estonia}} 

\author{
J. E. Lindberg\affil{Nordic Optical Telescope, Apdo. de Correos 474,
  38700 Santa Cruz de la Palma, Spain, and 
Centre for Star and Planet Formation, Natural History Museum of
Denmark, University of Copenhagen, \O ster Voldgade 5-7, DK-1350
Copenhagen, Denmark
}} 

\author{
T. Pursimo\affil{Nordic Optical Telescope, Apdo. de Correos 474,
  38700 Santa Cruz de la Palma, Spain}} 
 
\author{
P. Rodr\'\i guez-Gil\affil{Isaac Newton Group of Telescopes, 
Apdo. de Correos 321, 
E-38700 Santa Cruz de la Palma, 
Canary Islands, 
Spain, and Instituto de Astrof\'{\i}sica de Canarias, c/V\'{\i}a
L\'actea s/n, 38200 La Laguna, Tenerife, Spain}} 

\and

\author{
O. Vaduvescu\affil{Isaac Newton Group of Telescopes, 
Apdo. de Correos 321, 
E-38700 Santa Cruz de la Palma, 
Canary Islands
Spain, and Instituto de Astrof\'{\i}sica de Canarias, c/V\'{\i}a
L\'actea s/n, 38200 La Laguna, Tenerife, Spain}}


\begin{abstract}

The dust ejecta of Main-Belt Comet P/2010 A2 (LINEAR) have been 
observed with several telescopes at the 
at the Observatorio del Roque de los Muchachos 
on La Palma, Spain. Application 
of an inverse dust 
tail Monte Carlo method to the images of the 
dust ejecta from the object indicates
that a sustained, likely water-ice driven, activity over some 
eight months is the mechanism 
responsible for the formation of the observed
tail. The total amount of dust released is estimated to be 
5$\times$10$^7$ kg, which represents about 0.3\% of the nucleus mass. While 
the event could have been triggered by a 
collision, this cannot be decided from the currently available data.         

\end{abstract}


\section{Introduction}

On January 6, 2010, P/2010 A2 (LINEAR), a comet-like object, 
was discovered by the LINEAR sky survey. The orbital elements of the 
object ($a$ = 2.29 AU, $e$ = 0.12 and $i$ = 5.26 $\deg$) are typical 
of an inner Main-Belt asteroid belonging to the Flora collisional family, 
and suggest that it is unlikely to have originated in the classical
comet source regions (i.e., the Kuiper Belt or the Oort cloud). The
object reached perihelion, at a heliocentric distance of 
$r$=2.0 AU, on 4 December 2009.

Observations taken a 
few days after its discovery showed an inactive nucleus lying outside
a dust cloud that looks like a cometary tail without central
condensation \citep{Licandro10a,Licandro10b,Jewitt10}. This suggested that 
the observed dust cloud and the nucleus were the result of an impact 
between two previously unknown asteroids: the dust cloud is a plume 
of dust and the nucleus is what remains from the largest of the asteroids 
that collided.

Owing to its cometary-like aspect and orbital parameters, P/2010 A2 can 
be classified as a Main-Belt Comet \citep[MBC,][]{Hsieh06}.  In
contrast with the other known MBCs, however, P/2010 A2 has a significantly
smaller semi-major axis (2.29 AU versus 2.7 AU and $\sim$3.2 AU). 
Until now, the 
activity observed in MBCs was found to be compatible with a water-ice driven 
activation mechanism suggesting that those asteroids retained ice
layers  
below their surface and, under certain conditions, become 
``activated asteroids''. The presence of water-ice on the surface 
of 24 Themis, the parent asteroid of the Themis family MBCs 
\citep{Campins10,RivkinEmery10} strongly support this.

  One of the activation mechanisms could be a
collision between two asteroids. Collisions are known to take place 
regularly, but they are so rare that 
none of the dust plumes that they should generate has ever been seen. If 
P/2010 A2 is the debris of a collisional event, this would
be the first time that the ejecta from such a collision is observed 
soon after it happened. It would provide a unique opportunity to
learn something about asteroid collision processes and about the internal
composition of asteroids. Alternatively, if the observed activity is 
sustained in time as in comets, this leaves open two interesting 
problems: (1) if water-ice sublimation is the activation mechanism,
how does water-ice survive in an asteroid with such a small semi-major 
axis; or (2) is there any other mechanism capable of ejecting dust in a 
similar manner as water-ice sublimation does ?

In this paper we present and analyze images of P/2010 A2 obtained with
three 
telescopes at the Roque de los Muchachos Observatory (ORM), La Palma, 
Spain. In Sect. 2 the observations and data 
reduction are presented. In Sect. 3  we analyze the combined images
using the inverse Monte Carlo dust tail fitting method 
\citep[e.g.][]{Moreno04, Moreno09} to study possible ejection
scenarios. The conclusions are presented in Sect. 4.

\section{Observations and data reduction}

Images of P/2010 A2 were obtained in January 2010 with the 
Optical System for Imaging and low Resolution Integrated Spectroscopy
\citep[OSIRIS;][]{Cepa00, Cepa10} camera-spectrograph at the Gran
Telescopio Canarias (GTC), with the Auxiliary camera-spectrograph
(ACAM) at the William Herschel Telescope (WHT), and with the 
Andalucia
Faint Object Spectrograph and Camera (ALFOSC) at the Nordic Optical
Telescope (NOT) all located at the ORM.

The observational circumstances
are given in Table 1. For the observations with the NOT, the 
telescope was set at the comet's rate motion, while sidereal 
tracking was used for both WHT and GTC. In all cases the images were 
obtained in service mode by
telescope staff and were reduced by subtracting the overscan level and
flat fielded using standard procedures.

The observations with OSIRIS at the GTC  were made with the Sloan
g$^\prime$,r$^\prime$, 
and i$^\prime$ filters. OSIRIS provides a field of view 
7.8$\arcmin\times$7.8$\arcmin$ with a gap
of 9.2$\arcsec$ in the middle and a pixel scale of 
0.125$\arcsec$pixel$^{-1}$. To increase the signal-to-noise the data
were binned in 2$\times$2 pixels. The images 
were calibrated using photometric zeropoints determined from 
standard star observations.

The observations with ACAM at the WHT were made with the Sloan 
g$^\prime$ and r$^\prime$ filters. ACAM is mounted permanently 
at a folded-Cassegrain focus of the  telescope 
and has a circular field of view of 8$\arcmin$ diameter with a pixel
scale of 0.25 $\arcsec$pixel$^{-1}$. The 
resulting
calibrated WHT images show brightness levels in agreement with those
obtained at GTC within 10\%. 

The observations with ALFOSC at the NOT were made with 
standard Johnson-Cousins $V$ and $R$ filters. ALFOSC provides 
a field
of view of 6.5$\arcmin\times$6.5$\arcmin$ with a pixel scale of 0.19
$\arcsec$pixel$^{-1}$. These images were calibrated with stars in the
field of view using magnitudes from the 
USNO-B1.0 catalog \citep{Monet03}, which provides a photometric
accuracy of $\sim$0.3 mag.

Using the transformation equations of \cite{Fugukita96} and the
magnitude of the Sun in the standard Johnson-Cousins $V$ filter
\citep[$V_\sun$=--26.75,][]{Cox00}, we derive $r^\prime_\sun$=--26.96. Since
$R_\sun$=--27.29, the $r^\prime$ images of the object 
obtained at GTC and WHT were transformed to $R$
standard Johnson-Cousins magnitudes by subtracting 0.33 mag, 
where we assume for the object 
the same spectral dependence as for the Sun within the
bandpasses of these two red filters. The resulting GTC, WHT, and 
NOT $R$ magnitudes are consistent with their errors and we conclude
that the object did not change brightness significantly 
during our observations.

The available images in each night were 
combined in order to improve the signal-to-noise ratio and converted
to solar disk intensity units appropriate for the
analysis in terms of dust tail models. Figure 1 depicts some
representative images obtained on the different dates 
with the instruments mentioned above. In this figure one can 
clearly
see the inactive nucleus (marked with an arrow)  
lying outside the dust cloud that looks 
like a cometary tail. The absence of a dust cloud surrounding the
nucleus indicates that the dust emission has stopped before the
observations. Using aperture photometry, we determined the nucleus
$R$ magnitudes for the GTC, WHT, and NOT images whenever possible 
(see Table 1). The
aperture size was of 1.5$\arcsec$, and the nucleus profiles were 
stellar-like. For the 
GTC images, which have the best S/N ratio, and using the  
formalism by \cite{Bowell89} with a slope parameter of $G$=0.15, 
we obtained a nucleus absolute 
magnitude of $H$=21.3$\pm$0.3. Hence, a nucleus 
diameter of $D$=220$\pm$40 m can be determined,  
assuming a bulk albedo of $p$=0.11,
typical of a S-type asteroid. This assumption is based on the fact
that the 
S-type asteroids are the most common objects in the inner Main
Asteroid Belt \citep[e.g.][]{Bus02}.

The combined images from GTC and WHT obtained on the 17th and 21st
January show the best S/N ratio (see Fig. 1), and were selected for a 
more detailed analysis. To carry out the analysis in terms of dust
tail models an additional rebinning was made of the
images. The WHT image was
rebinned to 4 pixels, giving a spatial resolution of 772.4 km
pixel$^{-1}$, while the GTC image was rebinned to 3 pixels to give a 
resolution of 576.3 km pixel$^{-1}$. Finally, the rebinned images were
rotated to the $(N,M)$ coordinate system, where $M$ is the extended
radius vector from the Sun, and $N$ is perpendicular to $M$ 
and directed opposite to the object's motion along its orbit 
\citep{Fin68}.

\section{Application of the inverse dust tail model to P/2010 A2}

We have performed an analysis of both the WHT and GTC images by
the inverse Monte Carlo dust tail fitting code \citep[e.g.][]{Moreno04,
  Moreno09}, 
which is based on the same procedures as that
developed by Fulle 
\citep[e.g.][]{Fulle89, Fulle04}. Briefly, a 
large amount of particles (typically
10$^7$) of a selected 
size range are
released from the surface of the object in a given time interval,  
with an assumed velocity law, which may be a function of time and
particle size. The
particles are then submitted to the radiation pressure and gravity 
field of
the Sun, and their Keplerian orbits are computed. 
The ratio of the force exerted by the solar radiation 
pressure and the solar gravity 
is given by $\beta$, which can be expressed as $\beta =
C_{pr}Q_{pr}/(2\rho r)$, where $C_{pr}$=1.19$\times$ 10$^{-3}$ kg
m$^{-2}$, and $\rho$ is the particle density. For particle radii $r >$
0.25 $\mu$m, the radiation pressure coefficient is $Q_{pr}
\sim$ 1 \citep{Burns79}. The inverse method involves the inversion of the
overdetermined system of equations $AF=I$, where $A$ is the kernel
matrix containing the dust tail model, i.e., the surface density 
of the sample of particles integrated over time and $\beta$, $F$ is the
output vector which contains the time-dependent distribution of 
$\beta$, 
and $I$ is the brightness of the input image in the selected region of
the photographic $(N,M)$-plane.

The physical appearance of the tail, detached from the nucleus and
sun-ward oriented (see Figures 1 and 2), suggests that most of the 
dust was already ejected sometime  
well before the date of the observation,  
and should be composed of relatively large particles, otherwise they
would have been blown away by radiation pressure. This is clearly 
demonstrated in Figure 2, where the combined WHT image is shown along with the
synchrones \citep{Fin68} corresponding to ejection times from 272 days
before perihelion to 3 days after perihelion. Assuming a 
particle density
of 1000 kg m$^{-3}$, these synchrones would correspond
to particle sizes between 0.001 cm and 1 cm. To start the simulations,
and based on the synchrone map, we 
assumed initially that dust has been
ejected at any time starting approximately a year before the 
observations, with sizes in the range 0.001 cm to 1 cm. We fixed 
the particle density to $\rho$= 1000 kg m$^{-3}$,
and the albedo times the phase function to be $A_p$=0.04, i.e., a
Halley-like value. In 
principle, we assumed isotropic ejection. The better spatial 
resolution of the GTC image compared to the WHT image 
might suggest anisotropic ejection, which is   
particularly indicated by the shape of the isophotes near 
the nucleus. We will come back to this point later.

For lack of better information, we assumed an ejection
velocity as given by $v \propto \beta^{1/2}$, which from
hydrodynamical considerations is typical from gas
drag by sublimation processes. We did not assume any
temporal variation in the ejection velocities, which is reasonable
taking into account the low eccentricity of the orbit. We
found that the best fits to the images are found when the velocity is given by
$v=1100 \beta^{1/2}$ cm s$^{-1}$. This imply a velocity of 
8.5 cm s$^{-1}$ 
for a grain of $r$=1 cm. Using the formulae by \cite{Whipple51} for a 
nucleus of 220$\pm$40 m in diameter as given above and 
heating efficiency factor of 0.1 we get, at a heliocentric distance
of 2 AU, a range of velocities of 12--14 cm s$^{-1}$ 
(and smaller values for lower values of
the heating efficiency), which are not far from   
our model results. This would then be consistent with a typical 
cometary activity scenario where the outgassing is 
driven by water-ice sublimation. 

The result of the model fit to the WHT image is shown in
figure 3(a), and the derived 
dust mass loss rates as a function of time are shown in figure 4. The 
model  
isophotes reproduce quite closely those of the WHT image. The variation 
of the mass loss rate indicates that 
the onset of the activity occurred in late March 2009, 
with a maximum of 
activity around early June 2009, with a peak of dust loss rate of about 
5 kg s$^{-1}$, and 
decreasing to 0.1 kg s$^{-1}$ near perihelion in early December 2009. The
integrated dust loss is 5.4$\times$10$^7$ kg. Assuming a
spherically-shaped nucleus of diameter of 220 m, and a 
bulk density of 3000 kg m$^{-3}$,
the dust mass in the tail corresponds to 0.3\% of the total nucleus 
mass.  

We tried other model combinations to search for other
possible solutions. In particular,
we tried to fit the data by assuming  narrower time intervals for the
dust ejection, which could support an interpretation of a single
collision event without any further outgassing activity, but we could
not find any acceptable solutions even modifying also the particle sizes
and ejection velocities with values well above and below the expected
values that occur in the laboratory experiments such as presented by 
\cite{Nakamura91}.

If the isotropic ejection model with the same input parameters is
applied to the GTC image, we obtain the result shown in figure 3(b). In
this case, due to the fact that a certain portion of the tail 
is clipped, the dust mass loss differs from that found for the WHT
image, particularly in that the mass loss becomes zero at times earlier than
180 days pre-perihelion (see figure 4). For the same reason, the 
integrated ejected dust mass is also
lower at 1.7$\times$10$^7$ kg. Looking closely at the fitted
isophotes, some mismatch in the tail head near the
nucleus can be seen, and
also in the outermost isophote towards the sun-ward region. Therefore,
we 
decided 
to incorporate ejection anisotropy in the model by considering
emission from a selected active area at  
a particular region of a spherically symmetric rotating
nucleus. The rotation period was set to 3 hours. 
After trying tens of models with varying rotational parameters and 
location of the active area, a 
very good fit was obtained by taken the argument of the sub-solar
meridian at perihelion as $\Phi$=180$^\circ$, and the obliquity as 
$I$=95$^\circ$, with an active area occupying a part of the southern hemisphere
of the object in the latitude range [--90$^\circ$,--40$^\circ$]. The
resulting fit to the GTC observation is  
shown in figure 3(c). If a collision
triggered the outgassing event, it could have occurred at high
southern latitudes on the object. Although 
this model seems highly speculative, 
and other alternative scenarios exist, it does produce  
an excellent fit to the data. 

For the resulting time-averaged size distribution, a 
fit indicates a power-law with an index of --3.4$\pm$0.3,  a
value typically found in the analysis of dust ejected from 
comets \citep[e.g.][]{Fulle04}.   

The above fits indicate that 
the observations cannot be explained by dust released in a sudden
event such as a collision 
and subsequent ejection of asteroid surface material, but it 
does need a sustained cometary-like activity during several months 
after the onset of the activity as described above. Whether the onset
of the activity was triggered by a collision event or by another
mechanism that might be similar to those occurring in other
Main-Belt Comets \citep{Hsieh06}, can not be decided from the 
available data.

\section{Conclusions}

The application of the inverse dust tail Monte Carlo fitting method to
images of the tail of P/2010 A2 tail has revealed that the 
observed brightness distribution is only consistent with a 
sustained (possibly water-ice driven)  
cometary-like activity that spanned approximately  
8 months, which is 
estimated to have started in late March 2009. The
total amount of the ejected mass was about 5$\times$10$^7$
kg, when assuming an albedo times phase function of 4\%, and a particle density
of 1000 kg m$^{-3}$. It is not possible to asses   
whether the outgassing event was triggered 
by a collision or by another unspecified 
mechanism. However, the detection of a sustained cometary-like 
activity over a considerable period of time has 
obvious consequences for the internal structure of the object. It
might indicate that some subsurface ice layer exists in this
object. Another 
option is that there exists another mechanism which is capable 
of causing ejection of dust in a manner similar to water-ice
sublimation.   

A more sophisticated (but also
more speculative) dust tail model indicates that the dust ejection 
was localized and 
originated on the southern hemisphere of the nucleus, with 
rotation parameters 
$\Phi$=180$^\circ$, obliquity $I$=95$^\circ$, and a rotation period of $\sim$3
hours. These parameters can be confirmed with further follow-up observations.

\acknowledgments

We are grateful to an anonymous referee for his/her comments and
suggestions that help to improve the manuscript.

This work is 
based on observations made with the Gran Telescopio Canarias (GTC),
the William Herschel Telescope (WHT), and the Nordic Optical Telescope
(NOT), 
installed in the Spanish Observatorio del Roque de los Muchachos of
the Instituto de Astrof\'\i sica de Canarias, in the island of La
Palma. 

The William Herschel Telescope is operated 
by the Isaac Newton Group, and run by the Royal Greenwich 
Observatory at the Spanish Roque de los Muchachos Observatory in La Palma.

The Nordic Optical Telescope is operated on the island of La Palma jointly by Denmark, Finland, Iceland,
Norway, and Sweden, in the Spanish Observatorio del Roque de los
Muchachos of the Instituto de Astrofisica de Canarias.

The instrument ALFOSC on the NOT is owned 
by the Instituto de Astrof\'\i sica de Andaluc\'\i a (IAA) and operated at 
the Nordic Optical Telescope under agreement between IAA and the 
NBIfAFG of the Astronomical Observatory of Copenhagen.
 
This work was
supported by contracts AYA2007-63670, AYA2008-01720E, AYA2009-08190, and
FQM-4555 (Proyecto de Excelencia, Junta de Andaluc\'\i a). 
J. Licandro 
gratefully acknowledges support from the spanish ``Ministerio de
Ciencia 
e Innovaci\'on'' project AYA2008-06202-C03-02. 
J. L. Ortiz  
gratefully acknowledges support from the spanish ``Ministerio de 
Ciencia e Innovaci\'on'' project AYA2008-06202-C03-01.

\clearpage

\begin{figure}[ht]
\centerline{\includegraphics[scale=0.8,angle=-90]{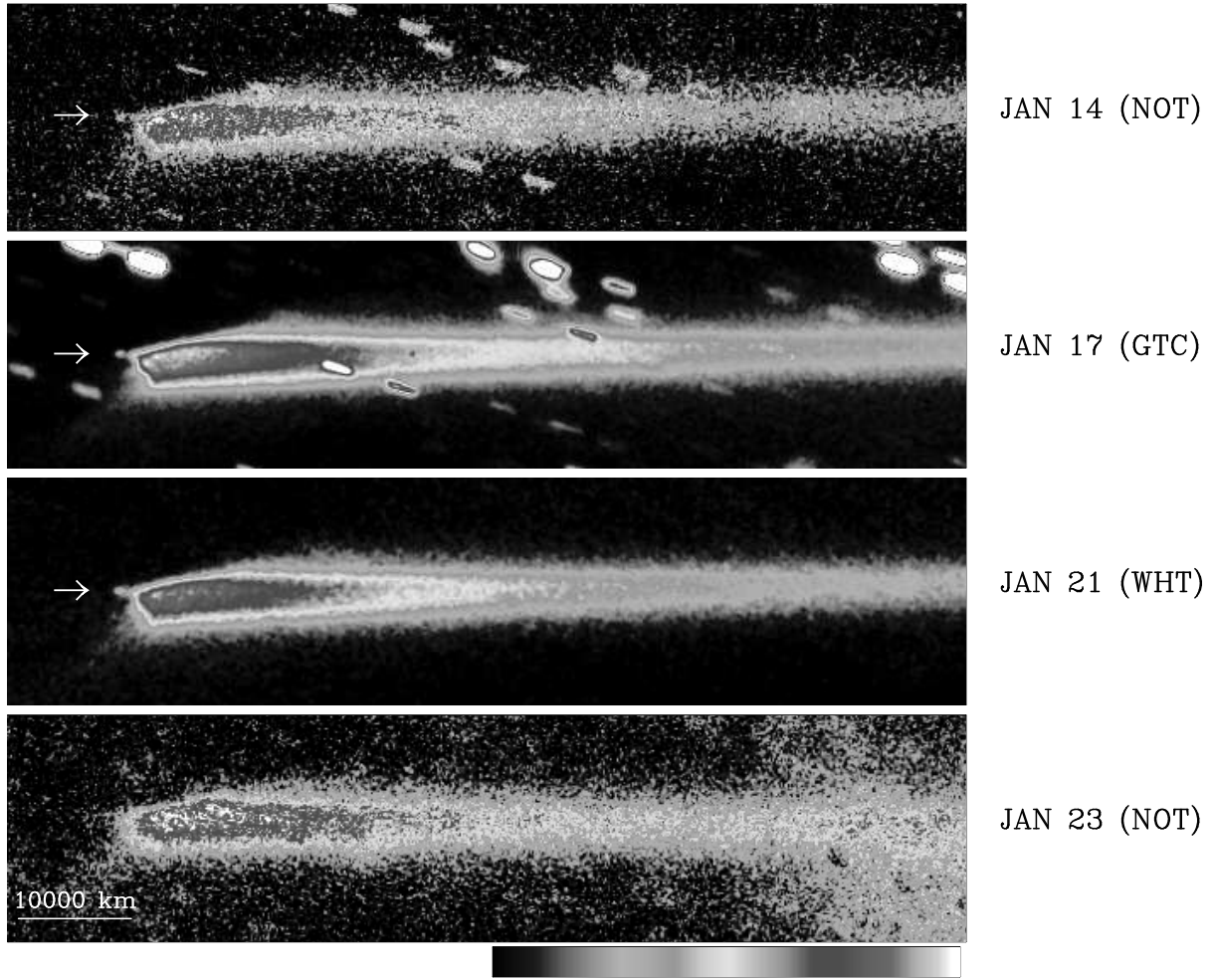}}
\caption{Images of P/2010 A2 (LINEAR) obtained at different dates with
  different telescopes at the Observatorio del Roque de los
  Muchachos in La Palma, as described in the text.  The telescope
  abbreviations are NOT = Nordic Optical Telescope, WHT = William
  Herschel 
Telescope, and GTC = Gran Telescopio Canarias. The images have
  been rotated in order to have the tail along the X-axis. The Sun
  direction is approximately toward +X. Whenever the nucleus is
  visible, it is marked with an arrow.  
 \label{fig0}}
\end{figure}

\clearpage

\begin{figure}[ht]
\centerline{\includegraphics[scale=0.8,angle=-90]{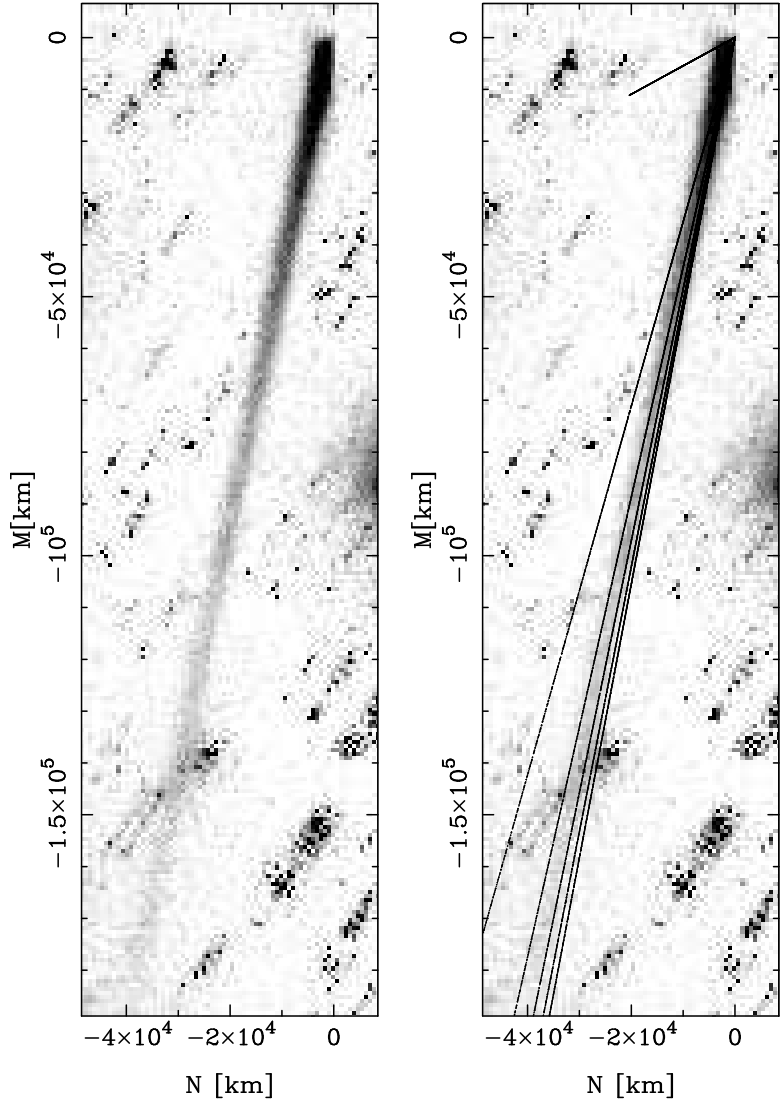}}
\caption{Left: Resulting calibrated and rebinned P/2010 A2 WHT image on the
  $(N,M)$ plane, acquired on January 21st, 2010. On the right, the
  synchrone map is overplotted. Synchrones are shown at (in
  counterclockwise sense)  +3, --52,
  --107, --162, --217, and --272 days since perihelion date, with
  maximum $\beta$=0.0595 (corresponding to a minimum particle radius of
  0.001 cm for a density of 1000 kg m$^{-3}$). \label{fig1}}
\end{figure}

\clearpage

\begin{figure}
\centerline{\includegraphics[scale=0.8,angle=-90]{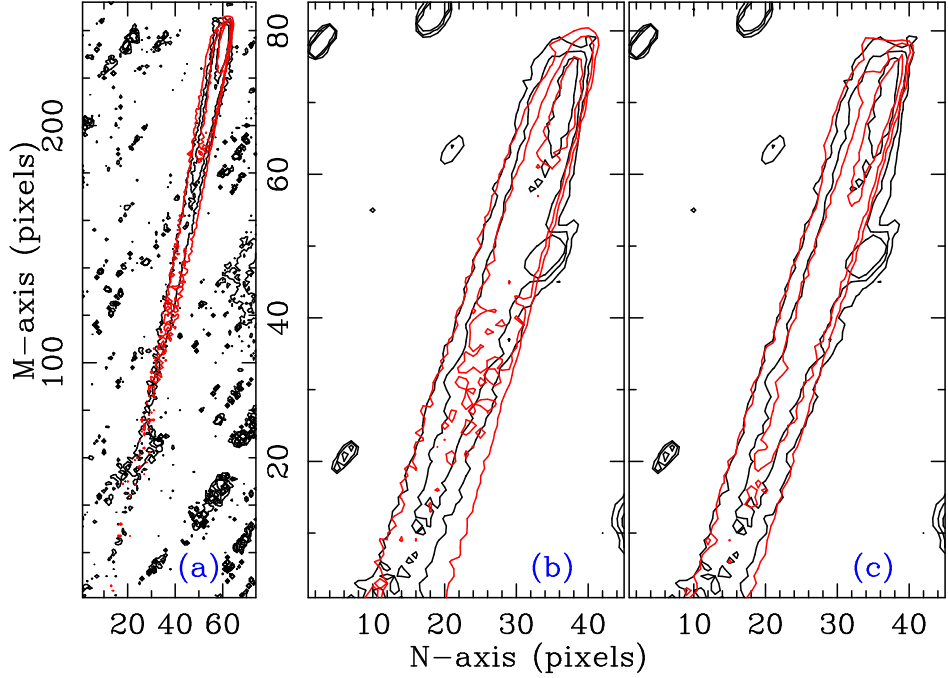}}
\caption{Results of the dust model applied to the WHT and GTC
  images. Coordinate axes correspond to the $(N,M)$ system
  (see text). Black contours, with isophote levels of 
5$\times$10$^{-15}$, 10$^{-14}$, and
  2$\times$10$^{-14}$ solar disk intensity units, 
correspond to the observations. Red
  contours correspond to the model. 
Panel (a): The isotropic model applied to the 
WHT image. The physical
  dimensions of the images  
are 57154 km $\times$ 194632 km. Panel (b): The isotropic model applied 
to the GTC image.  The physical
  dimensions of the images  
are 25932 km $\times$ 48407 km. Panel (c): The anisotropic 
model applied to the GTC image, with same physical dimensions as 
  those of panel (b).  
\label{fig3}}
\end{figure}

\clearpage

\begin{figure}
\centerline{\includegraphics[scale=0.8,angle=-90]{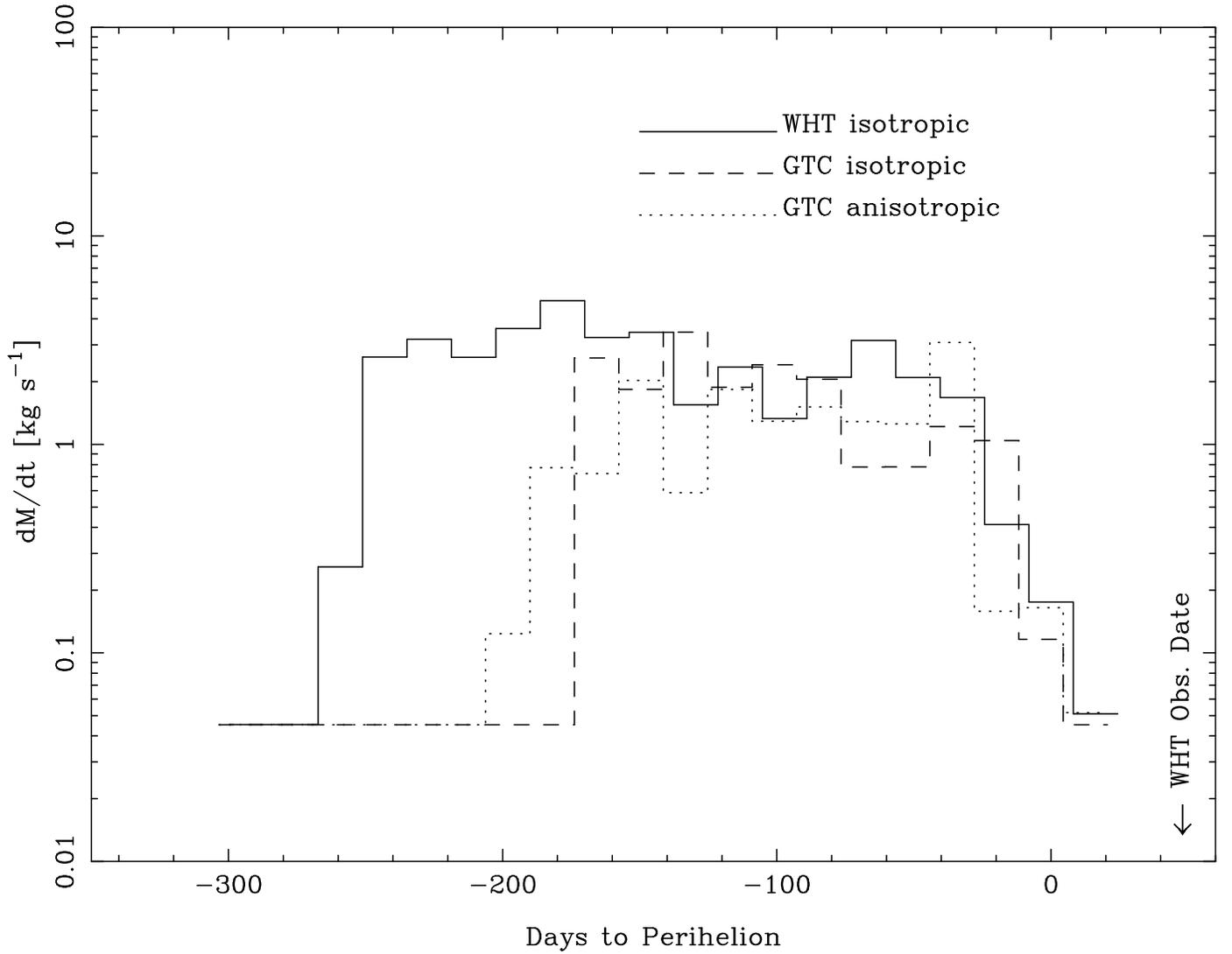}}
\caption{The derived dust mass loss rates versus time 
from the WHT and GTC images
  using isotropic and anisotropic dust ejection models as
  indicated. The arrow marks the date of the observation of the WHT image. 
\label{fig4}}
\end{figure}

\clearpage

\begin{deluxetable}{cccccc}
\tablewidth{0pt}
\tablecaption{Observational circumstances of P/2010 A2. Position data include
  heliocentric distance $r$, geocentric distance $\Delta$, and phase
  angle $\alpha$. Last column shows the Johnson-Cousins
  $R$-magnitude of the nucleus with their estimated uncertainties.}
\tablehead{
\colhead{UT date}  & Telescope & \colhead{$r$ (AU)} &
\colhead{$\Delta$ (AU)}& 
\colhead{$\alpha^\circ$} & \colhead{Nucleus $R$-mag}} 

\startdata

14 Jan 2010 & NOT 2.56 m & 2.013 & 1.039 & 5.53 & 23.3$\pm$0.4 \\  
17 Jan 2010 & GTC 10.4 m & 2.014 & 1.047 & 7.20 & 23.5$\pm$0.3 \\
21 Jan 2010 & WHT 4.2 m  & 2.016 & 1.061 & 9.32 & 23.2$\pm$0.3 \\
23 Jan 2010 & NOT 2.56 m & 2.016 & 1.065 & 9.87 &  -- \\

\enddata

\end{deluxetable}

\end{document}